\documentclass[12pt]{article}
\usepackage{amssymb}
\newcommand{\goes}{\rightarrow}
\newcommand{\PD}[2]{{\partial{#1}\over\partial{#2}}}
\newcommand{\NN}{\nonumber}
\newcommand{\bra}[1]{\langle{#1}|}
\newcommand{\ket}[1]{|{#1}\rangle}
\newcommand{\ssl}{\left[ }
\newcommand{\ssr}{\right] }
\newcommand{\expect}[1]{\langle{#1}\rangle}

\newcommand{\BE}{\begin{equation}}
\newcommand{\EE}{\end{equation}}
\newcommand{\BEA}{\begin{eqnarray}}
\newcommand{\EEA}{\end{eqnarray}}

\newcommand{\HEPTH}[1]{{\tt hep-th/{#1}}}
\newcommand{\IJMP}[3]{Int. J. Mod. Phys. {\bf #1}{(#2)}{#3}}
\newcommand{\JHEP}[3]{JHEP~{\bf #1}{(#2)}{#3}}

\newcommand{\MPL}[3]{Mod. Phys. Lett. {\bf #1}{(#2)}{#3}}
\newcommand{\NP}[3]{Nucl. Phys. {\bf #1}{(#2)}{#3}}
\newcommand{\PL}[3]{Phys. Lett. {\bf #1}{(#2)}{#3}}
\newcommand{\PR}[3]{Phys. Rev. {\bf #1}{(#2)}{#3}}
\newcommand{\PRL}[3]{Phys. Rev. Lett. {\bf #1}{(#2)}{#3}}

\def\ap{\alpha^{\prime}}
\def\at{\tilde{\alpha}}
\def\p{\partial}
\def\pb{\bar{\partial}}

\def\zb{\bar{z}}
\def\wb{\bar{w}}

\def\12{\frac{1}{2}}
\def\bea{\begin{eqnarray}}
\def\eea{\end{eqnarray}}
\def\ba{\begin{array}}
\def\ea{\end{array}}

\def\one-loop{\mbox{\scriptsize one-loop}}

\def\psib{\bar{\psi}}

\def\G{\Gamma}

\catcode`\@=11

\def\@normalsize{\@setsize\normalsize{15pt}\xiipt\@xiipt
\abovedisplayskip 14pt plus3pt minus3pt%
\belowdisplayskip \abovedisplayskip
\abovedisplayshortskip  \z@ plus3pt%
\belowdisplayshortskip  7pt plus3.5pt minus0pt}

\def\small{\@setsize\small{13.6pt}\xipt\@xipt
\abovedisplayskip 13pt plus3pt minus3pt%
\belowdisplayskip \abovedisplayskip
\abovedisplayshortskip  \z@ plus3pt%
\belowdisplayshortskip  7pt plus3.5pt minus0pt
\def\@listi{\parsep 4.5pt plus 2pt minus 1pt
            \itemsep \parsep
            \topsep 9pt plus 3pt minus 3pt}}

\def\underline#1{\relax\ifmmode\@@underline#1\else
        $\@@underline{\hbox{#1}}$\relax\fi}
\@twosidetrue

\relax

\catcode`@=12

\catcode`\@=11

\def\section{\@startsection{section}{1}{\z@}{3.5ex plus 1ex minus
   .2ex}{2.3ex plus .2ex}{\large\bf}}

\def\thesection{\Roman{section}.}

\def\appendix{\setcounter{section}{0}
        \def\thesection{Appendix }
        \def\theequation{\Alph{section}.\arabic{equation}}}

\def\ps@headings{\def\@oddfoot{}\def\@evenfoot{}
\def\@oddhead{\hbox{}\hfill
        \makebox[.5\textwidth]{\raggedright\ignorespaces --\thepage{}--
        \hfill {}}}
\def\@oddhead{\hbox{}\hfill --\thepage{}-- \hfill
        {}}
\def\@evenhead{\@oddhead}
\def\subsectionmark##1{\markboth{##1}{}}
}

\ps@headings

\catcode`\@=12

\relax

\def\figcap{\section*{Figure Captions\markboth
        {FIGURECAPTIONS}{FIGURECAPTIONS}}\list
        {Fig. \arabic{enumi}:\hfill}{\settowidth\labelwidth{Fig. 999:}
        \leftmargin\labelwidth
        \advance\leftmargin\labelsep\usecounter{enumi}}}
 \relax
\def\tablecap{\section*{Table Captions\markboth
        {TABLECAPTIONS}{TABLECAPTIONS}}\list
        {Table \arabic{enumi}:\hfill}{\settowidth\labelwidth{Table 999:}
        \leftmargin\labelwidth
        \advance\leftmargin\labelsep\usecounter{enumi}}}
 \relax
\def\reflist{\section*{References\markboth
        {REFLIST}{REFLIST}}\list
        {[\arabic{enumi}]\hfill}{\settowidth\labelwidth{[999]}
        \leftmargin\labelwidth
        \advance\leftmargin\labelsep\usecounter{enumi}}}
 \relax

\catcode`\@=11

\def\ps@headings{\def\@oddfoot{}\def\@evenfoot{}
\def\@oddhead{\hbox{}\hfill
        \makebox[.5\textwidth]{\raggedright\ignorespaces --\thepage{}--
        \hfill {}}}
\def\@evenhead{\@oddhead}
\def\subsectionmark##1{\markboth{##1}{}}
}

\ps@headings

\relax

\newskip\humongous \humongous=0pt plus 1000pt minus 1000pt

\newif\ifdtup

\def\bra#1{\left\langle #1\right|}
\def\ket#1{\left| #1\right\rangle}

\def\beq{\begin{equation}}
\def\eeq{\end{equation}}

\def\beqn{\begin{eqnarray}}
\def\eeqn{\end{eqnarray}}
\relax

\def\G2{{\; \rm GeV/}c^2}
\def\G{\; \rm GeV}

\def\dotx{\dotx{\dot\overline{x}}}

\relax

\def\p{\partial}

\begin{document}

%
%
\begin{titlepage}

\renewcommand{\thefootnote}{\fnsymbol{footnote}}

\begin{flushright}
      \normalsize
      May, 2001  \\     
     OU-HET 387, \\ 
     YITP-01-46, \\
         hep-th/0105247  \\
\end{flushright}

%
\begin{center}
  {\large\bf  Some Computation on $g$ Function and Disc  
  Partition Function and Boundary String Field Theory }
\end{center}

\vfill
\begin{center}
    {  {A. Fujii${}^1$}\footnote{e-mail address:
                     fujii@yukawa.kyoto-u.ac.jp} and
{ H. Itoyama${}^2$}\footnote{e-mail address:
                       itoyama@het.phys.sci.osaka-u.ac.jp}
       }\\
\end{center}

\vfill

\begin{center}
    ${}^1$\it Yukawa Institute for Theoretical Physics,
       Kyoto University, \\
       Kyoto 606-8502, Japan \\
~\\

    ${}^2$\it  Department of Physics,
        Graduate School of Science, Osaka University,\\
        Toyonaka, Osaka 560-0043, Japan
\end{center}

\vfill


\begin{abstract}

Quadratic tachyon profile has been  discussed 
in the boundary string field theory.
We here compute  the $g$-function by 
factorizing the cylinder amplitude. 
The answer is compared with the disc partition function.  
The boundary state is constructed.  
We extend these computations to those of the  
boundary sine-Gordon model at 
the free fermion point.

\end{abstract}

\vfill

\setcounter{footnote}{0}
\renewcommand{\thefootnote}{\arabic{footnote}}

\end{titlepage}


The boundary string field (BSFT) theory\cite{witten} has been 
recently adopted for the analysis of Sen's 
tachyon condensation\cite{sen} both in bosonic 
and supersymmetric string theory. 
In the bosonic BSFT the spacetime string action $S$  
is conjectured to satisfy
\cite{shatashvili}-\cite{tseytlin}
\BE
\PD{S}{\lambda_i}={\cal G}^{ij}{\hat \beta}_j,
\label{eq:dSdl}
\EE
where $\lambda_i$ are the boundary coupling constants, 
${\cal G}^{ij}$ is their metric, and ${\hat \beta}_i$ are  
the beta functions. This differential equation 
for $S$ has the same form as that conjectured 
for the ground 
state degeneracy ($g$-function)\cite{affleck-ludwig} 
in statistical models. 
In fact the studies based on this similarity 
have been progressed by several authors
\cite{harvey-kutasov-martinec}\cite{tensionisdimension}\cite{fi1}. 
  
In this letter we consider the model in which the open 
string on-shell action is deformed at the boundary 
by the quadratic term in $X$\cite{witten}. 
This model has been used in the evaluation 
of the exact tachyon potential
\cite{gerasimov-shatashvili}-\cite{tseytlin} 
and can be obtained 
as a limit of boundary sine-Gordon (BSG) model. 
The Lagrangian of the BSG model is
\BEA
{\cal S}&=&{1\over 4\pi\ap}
\int_{\Sigma}\sqrt{{\rm det}\,\eta}\,\eta^{ij}\,\p_i X 
\p_j X\,d^2\sigma+\int_{\p\Sigma}T(X)\,dy,\NN\\
T(X)&=&\zeta 
\cos\left({2\pi\over R}X\right),
\label{eq:bsg}
\EEA
where $\Sigma$ is a worldsheet of open string, 
$\eta_{ij}$ is its metric, and $\p\Sigma$ is the 
boundary parameterized by $y$. 
If the quantity $\lambda = R^2 /4\pi^2\ap -1$ is a 
non-negative integer, the model is exactly solvable
\cite{bsG} 
and $\lambda$ corresponds to the number of the species 
of the breathers. If we take the limit $R\goes\infty$ 
and let the parameter $\zeta$ scale appropriately, 
$T(X)$ becomes a quadratic function of $X$, namely, 
$T(X)=a+u X^2$. Let us call this the boundary 
quadratic deformation (BQD) model.   

We evaluate the partition functions of the BQD model 
on a cylinder and a disc. The disc partition function 
followed by the spacetime string action is already 
evaluated in ref.\cite{witten}. 
Our main interest in this letter goes to the 
$g$-description. Thus we first calculate 
the cylinder partition function and extract $g$ 
by means of the finite size scaling.  
Because the BQD is a specific case of the 
BSG model with $R\goes\infty$ limit, the 
direct calculation of the functional determinant is 
more suitable than the thermodynamic Bethe ansatz (TBA) 
method\cite{fendley-saleur-warner}
\cite{leclair-mussardo-saleur-skorik}, 
which includes breathers of 
an infinite number of species. 

Secondly we construct the boundary state\cite{Callan1}  
on a disk which reproduces the known partition 
function\cite{witten} and 
can be used to define the $g$-function on the disk. 
We will see that $S$ as a function of the 
boundary coupling constant on the disc has a pole  
at every minus integer whereas $g$ on the cylinder 
has a zero there. 

Finally we consider the BSG model at the 
free fermion (FF) point, $4\pi^2 \ap/R^2 =1/2$. The 
$g$-function of the FF model on a cylinder 
is obtained in 
refs.\cite{leclair-mussardo-saleur-skorik}\cite{chatterjee}\cite{konik}.  
We evaluate its partition function on a disc 
by means of the fermionization method. We find a 
similar result to the BQD case: $g$ on the cylinder 
has a zero at every  
minus half-integer while the partition function on 
the disc possesses a pole there.     

Let us consider the BQD model 
on a cylinder and evaluate its partition function. 
We remark that a similar analysis on an annulus can be 
found in ref.\cite{suyama}. 
Let the length of the cylinder be $l$ 
and the radius be $r$. The temperature is identified with 
$\theta=1/2\pi r$. The Lagrangian and the partition function 
of the BQD model are 
\BE
L={1\over 4\pi\ap}\left[
\int^{l}_{0}dx(\p_a X)^2 -vX(t,0)^2
\right]
\EE
and 
\BE
Z=\int{\cal D}X\,\exp\left(-\int^{2\pi r}_{0}L\,dt\right),
\EE
respectively. The boundary condition at $x=0$ is 
a mixed type: 
\BE
\p_{x}X+vX=0. \label{eq:mixed}
\EE
We impose the Dirichlet type condition at $x=l$ and 
the periodic boundary condition in $t$-direction.
\footnote{This boundary condition is similar to but 
distinct from that of $p-p'$ open string with $B$-field
\cite{SW}\cite{CIMM}.} 

Let us calculate the partition function in 
the Lagrange formalism. 
We expand the field $X$ as 
\BE
X(t,x)=\sum_{m\in\mathbb{Z}}\sum_{j}{\cal A}_{m,\mu_j}
e^{imt/r}\sin\mu_j (x-l),\quad 
{\cal A}_{m,\mu_j}={\cal A}_{-m,\mu_j}.
\EE
Because we assume the periodic boundary condition in 
$t$-direction, $m$ runs over integers. In $x$-direction, 
we impose the mixed boundary condition (\ref{eq:mixed}), 
which gives the equation determining $\mu_j$:
\BE
\mu_j=v\tan(\mu_j l). \label{eq:tan}
\EE
Therefore, the partition function becomes
\BE
Z'={\rm Det'}^{-1/2}(-\p_t^2 -\p_x^2 )=\left(
(\prod_{m=0}^{\infty}\prod_{j}^{})'
\left[
(m/r)^2+\mu_j^2
\right]
\right)^{-1/2}.\label{eq:partition-function}
\EE
The complete partition function $Z$ includes 
the contribution from the zero mode integration. 

Applying the standard zeta-function regularization, we obtain
\BE
Z'=\prod_{\mu_j >0}(e^{\pi r\mu_j}-e^{-\pi r\mu_j})^{-1}.
\EE
Let us define a function 
\BE
f(p)=p-{1\over l}\tan^{-1}{p\over v}.\label{eq:phase-function}
\EE
If a certain $\mu_j$ satisfies (\ref{eq:tan}), 
one can see $f(\mu_j)=\pi I_j /l$, where 
$I_j$ is an integer. If $vl>1$, $f(p)$ is a 
monotonically increasing function. Therefore, 
$I_{j+1}-I_{j}=1$. 
Let us introduce the density function 
\BE
\rho(\mu_j )=\lim_{l\goes\infty}{1\over l(\mu_{j+1}-\mu_j )}.
\EE
In the limit $l\goes\infty$, $\mu_j$ develops continuously 
in $(0,\infty)$. From the form of (\ref{eq:phase-function}), 
we see
\BE
\pi\rho(\mu)=1-{1\over vl}{1\over 1+(\mu/v)^2}+O(l^{-2}).
\EE
Hence, the free energy is represented like
\BEA
F'/\theta=-\ln Z'&=&\sum_{\mu_j>0}\left[ 
\pi r\mu_j +\ln(1-e^{-2\pi r\mu_j})
\right],\NN\\
&=&\int^{\infty}_{0}
\left[ 
\pi r\mu +\ln(1-e^{-2\pi r\mu})
\right] \,l\rho(\mu)d\mu.
\EEA
Let us recall the definition of $g$-factor in terms 
of the finite size scaling:
\BE
F=f_{0}l-\theta\ln g+O(l^{-1}).
\EE
We conclude 
\BE
\ln g = {v\over \pi}\int^{\infty}_{0}
{d\mu\over \mu^2+v^2}
\ln(1-e^{-2\pi r\mu})-\ln\sqrt{rv} 
=\ln\left[ {\sqrt{2\pi}\over\Gamma(u+1)}
\left({u\over e}\right)^u\right],
\label{eq:gbqd}
\EE
where $u=rv$. 
We have subtracted $\ln\sqrt{rv}$, taking  
the zero mode integration into account. 
In (\ref{eq:gbqd}), 
taking the limits $u\goes 0$ (UV limit) and 
$u\goes\infty$ (IR limit), we see 
$g_{\rm UV}/g_{\rm IR}=\infty$. 
This is consistent with the $g$-theorem\cite{affleck-ludwig}.  
We should remark here that from (\ref{eq:gbqd}) 
$g$ has zeros at $u=-1,-2,\cdots$.   

Let us consider BQD model on a disc ${\cal D}$: 
$0\leq r\leq c$, $0\leq\sigma\leq 2\pi$, 
where $c$ is the radius of ${\cal D}$.  
The action is 
\BE
{\cal S}={1\over 4\pi\ap}\int_{\cal D}\sqrt{{\rm det}\,\eta}\,
\eta^{ij}g_{\mu\nu}\p_i X^\mu \p_j X^\nu\,d^2\sigma
+{1\over 4\pi\ap}\int_{\p{\cal D}}
(a+u_{\mu\nu}X^\mu X^\nu)\,d\sigma,
\label{eq:diskaction}
\EE
where we set $\sigma_1 =r$ and $\sigma_2 =\sigma$. 
$\eta^{ij}$ and $g_{\mu\nu}$ are the metrics of the 
world sheet ${\cal D}$ and the spacetime, respectively, 
$a$ and $u_{\mu\nu}$ are the coupling constants at 
the boundary. 
From the action (\ref{eq:diskaction}), we should impose 
the boundary condition on $\p{\cal D}$ as 
\BE
\left.
g_{\mu\nu}r\PD{}{r}X^\nu +u_{\mu\nu}X^\nu
\right|_{r=c} =0. \label{eq:bc}
\EE

In ref.\cite{witten} 
the partition function of the BQD model is calculated 
on the disc with unit radius, {\it i.e.} $c=1$ with 
$g_{\mu\nu}=\delta_{\mu\nu}$ and 
$u_{\mu\nu}=u_\mu \delta_{\mu\nu}$. The result is 
\BE
Z=e^{-a}\prod_{\mu}Z_1 (u_\mu),\quad 
Z_1 (u)=\sqrt{u}e^{\gamma u}\Gamma(u). \label{eq:pfdisk}
\EE

Let us consider the partition function (\ref{eq:pfdisk}) 
in the closed string picture. For this sake, we introduce 
the boundary state satisfying
\BE
\bra{\cal{B}}\left.\left(
g_{\mu\nu}r\PD{}{r}X^\nu +u_{\mu\nu}X^\nu
\right)\right|_{r=c}=0. \label{eq:boundarystate}
\EE
To determine the explicit form of $\bra{\cal{B}}$, we utilize 
the mode expansion of $X$ in terms of the complex 
coordinates, $z=re^{i\sigma}$ and $\zb=re^{-i\sigma}$:
\BEA
X^\mu (r,\sigma)&=&X^\mu _{\rm R}(z)+X^\mu _{\rm L}(\zb),\NN\\
X^\mu _{\rm R}(z)&=&
{1\over 2}
X^\mu _0
-{i\ap\over 2}p^\mu
\ln z+i\sqrt{{\ap\over 2}}
\sum_{n\neq 0}\alpha^\mu _n {z^{-n}\over n}, \\
X^\mu _{\rm L}(\zb)&=&
{1\over 2}
X^\mu _0
-{i\ap\over 2}p^\mu
\ln \zb +i\sqrt{{\ap\over 2}}
\sum_{n\neq 0}\at^\mu _n {\zb^{-n}\over n}\NN
\EEA
with
\BE
\ssl X_0 ^\mu , p^\nu \ssr =ig^{\mu\nu},\quad
\ssl \alpha_m ^\mu , \alpha_n ^\nu\ssr =
mg^{\mu\nu}\delta_{m+n,0},\quad 
\ssl \at_m ^\mu , \at_n ^\nu\ssr =mg^{\mu\nu}\delta_{m+n,0},
\quad\ssl \alpha_m ^\mu , \at_n ^\nu\ssr =0.
\EE
With the above expression, let us seek for the boundary 
state of the following form 
\BE
\bra{\cal{B}}=
\bra{0}\exp\left(-{1\over 2}X_0^\mu A_{\mu\nu} X_0^\nu\right)
\exp\left(
\sum_{m=1}^{\infty}\at_m^\mu C_{\mu\nu}^{(m)}\alpha_m^\nu
\right), \label{eq:ansatzb}
\EE
where $\bra{0}$ is the out vacuum state:
\BE
\bra{0}\alpha_{-n}^\mu =\bra{0}\at_{-n}^\mu =0\quad{\rm for~}n>0.
\EE 
Upon setting $g_{\mu\nu}=\delta_{\mu\nu}$ and 
$u_{\mu\nu}=u_\mu \delta_{\mu\nu}$, 
we see that the boundary state  
(\ref{eq:ansatzb}) satisfies the condition 
(\ref{eq:boundarystate}) if the parameters are
\BE
A_{\mu\nu}={1\over\ap}{u_\mu\over 1+u_\mu\ln c}
\,\delta_{\mu\nu},
\quad 
C_{\mu\nu}^{(m)}=-{c^{-2m}\over m}
{m-u_\mu\over m+u_\mu}\,\delta_{\mu\nu}.
\label{eq:parabs}
\EE

This boundary state (\ref{eq:ansatzb}) with (\ref{eq:parabs}) can 
be used in the calculation of the correlation functions. 
For example, the two point function can be calculated as
\BEA
\!\!\!{2\over\ap}\expect{X^\mu (z,\zb)X^\nu (w,\wb)}&=&
{2\over\ap}
{\bra{\cal{B}}X^\mu (z,\zb)X^\nu (w,\wb)\ket{0}\over\expect{B|0}}\NN\\
&=&\!\!\!\!\!
-\delta^{\mu\nu}\ln(|z-w|/c)^2-\delta^{\mu\nu}
\ln|1-z\wb /c^2|^2\NN\\
&&\!\!\!\!\!\!+\delta^{\mu\nu}\cdot{2\over u_{\mu}}-2\delta^{\mu\nu}u_\mu\sum_{m=1}^{\infty}{1\over m(m+u_\mu )}
\ssl(z\wb /c^2)^m +(\zb w/c^2)^m \ssr\!\!,
\EEA
which agrees with the result in ref.\cite{witten} 
by means of Green's differential equation. 
Hence, the boundary state 
(\ref{eq:ansatzb}) with (\ref{eq:parabs}) 
reproduces the disc partition function 
(\ref{eq:pfdisk}) precisely. The partition function 
(\ref{eq:pfdisk}) and the spacetime action 
$S(=i_{W}Z)$ have poles at $u_\mu =-1,-2,\cdots$. 

Finally let us evaluate the partition function of 
the BSG on a disk at the FF point, namely, 
\BE
{\cal S}_{\rm SG}=
{1\over 8\pi}\int_D d^2\sigma \p X\pb X
+\zeta\int_{\p D}d\sigma\cos{1\over 2}X, 
\label{eq:bSG}
\EE
where we set $\ap=2$. 
To execute the calculation, it is useful to apply 
the description in terms of massless fermions
\cite{chatterjee}-\cite{luca-rafael} of the 
action (\ref{eq:bSG}). 
The fermionic action ${\cal S}$ consists of four terms, 
${\cal S}={\cal S}_{\rm free}^{(0)}+{\cal S}_{\rm free}^{(1)}
+{\cal S}_{\rm int}^{(0)}+{\cal S}_{\rm int}^{(1)}$, where  
\BE
{\cal S}_{\rm free}^{(0)}=
{1\over 8\pi}\int^{c}_{0}rdr\int^{2\pi}_{0}d\sigma
\ssl \psi_{+}\pb\psi_{-} +\psi_{-}\pb\psi_{+}
-\psib_{+}\p\psib_{-} -\psib_{-}\p\psib_{+}\ssr
\label{eq:action0}
\EE
is a free fermion action in the bulk, and 
the boundary terms are
\BEA
{\cal S}_{\rm free}^{(1)}&=&
-{c\over 8\pi}\int^{2\pi}_{0}d\sigma
\ssl \psi_+ \psib_+ +\psi_- \psib_- \ssr ,\\
{\cal S}_{\rm int}^{(0)}&=&
{1\over 2\pi}\int^{2\pi}_{0}d\sigma a_+ \p_\sigma a_- ,\\
{\cal S}_{\rm int}^{(1)}&=&
{c\zeta\over 32\pi^2}\int^{2\pi}_{0}d\sigma
\ssl
e^{i\sigma /2}(\psi_+ a_- +a_+ \psi_- )+
e^{-i\sigma/2}(\psib_- a_+ +a_- \psib_+ )
\ssr. \label{eq:action}
\EEA

We can obtain the equation of motion 
\BE
\pb\psi_{\pm}=\p\psib_{\pm}=0,
\EE
and the boundary conditions 
\BEA
a_+ &=& a_- = e^{i\sigma/2}\psi_+ -e^{-i\sigma/2}\psib_- 
= -e^{i\sigma/2}\psi_- +e^{-i\sigma/2}\psib_+ \label{eq:bc00}\\
0&=&\p_\sigma (e^{-i\sigma/2}\psib_+ -e^{i\sigma/2}\psi_- )
-B(e^{-i\sigma/2}\psib_- -e^{i\sigma/2}\psi_- ), \label{eq:bc01}
\EEA
where dimensionless parameter is introduced, 
$B=\zeta^2 c/64\pi^2$. 

Substituting (\ref{eq:bc00}) into (\ref{eq:action}), 
we obtain
\BE
-{d\over d\zeta}\ln Z=
{c\over 2}\int^{2\pi}_{0}d\sigma 
\ssl
e^{i\sigma/2}(\expect{\psi_+ a_- }+\expect{a_+ \psi_- })
+e^{-i\sigma/2}(\expect{\psib_- a_+ }+\expect{a_- \psib_+ })
\ssr\label{eq:diffZ}
\EE
With the help of Green's functions, we 
can calculate the partition function 
by integrating (\ref{eq:diffZ}). As a result, we should 
integrate
\BE
-{d\over d\zeta}\ln Z
=-\lim_{\epsilon\goes 0}
{1\over \pi\zeta}\int^{2\pi}_{0}d\sigma
\sum_{k=0}^{\infty}
{k+1/2\over k+1/2+B}
e^{-ik\epsilon}. \label{eq:diffZ1}
\EE 
The above quantity diverges as $\epsilon\goes 0$, and thus 
needs regularization. If we subtract 
$1-B/(k+1/2)$ in the summation, we obtain
\BE
-{d\over d\zeta}\ln Z_{\rm reg}
=-{B^2\over \pi\zeta}\int^{2\pi}_{0}d\sigma 
\sum_{k=0}^{\infty}
{1\over (k+1/2)(k+1/2+B)}. \label{eq:regZ}
\EE
We can integrate (\ref{eq:regZ}) to yield  
\BE
Z_{\rm reg} = {\Gamma(2B)\over\Gamma(B)}=
{2^{2B-1}\over\sqrt{\pi}}\Gamma(B+1/2),
\EE
which has poles at $B=-1/2,-3/2,\cdots$. 
On the other hand, the $g$-function defined on a cylinder is 
known\cite{leclair-mussardo-saleur-skorik}\cite{chatterjee}\cite{konik} as  
\BE
g={\sqrt{2\pi}\over\Gamma(\alpha+1/2)}
\left({\alpha\over e}\right)^\alpha,
\EE 
where $\alpha = 32\pi^2 r \zeta^2$.  

We finally point out again a curious reciprocal relation between 
the disc partition function and the $g$-function extracted from the 
cylinder partition function, which is common to the two examples considered 
in this letter. In the  analytic coupling constant plane, the disc partition 
function has integer spaced 
poles in the negative real axis whereas the $g$-function 
possesses zeros precisely at these points.

{\bf Acknowledgments}\\ 
The authors would like to thank T.~Nakatsu and T.~Suyama 
for helpful discussions.  
This work is supported in part by the Grant-in-Aid  
for Scientific Research No.12640272 from 
the Ministry of Education, Science and Culture, Japan. 
AF was partially supported by The Yukawa Memorial 
Foundation. 


\end{document}